\documentclass[aps,prb,preprint,groupedaddress]{revtex4-1}
\usepackage{graphicx}

\begin{document}


\title{
Deconfined criticality 
for the $S=1$ spin model
on the spatially anisotropic triangular lattice
}

\author{Yoshihiro Nishiyama} 
\affiliation{Department of Physics, Faculty of Science,
Okayama University, Okayama 700-8530, Japan}

\date{\today}

\begin{abstract}
The quantum $S=1$ spin model
on the 
spatially anisotropic 
triangular lattice is investigated numerically.
The nematic and valence-bond-solid (VBS) phases
are realized by adjusting the spatial anisotropy
and the biquadratic interaction.
The phase transition between the nematic and VBS phases 
is expected to be a continuous one with
unconventional critical indices
(deconfined criticality).
The geometrical character (spatial anisotropy)
is taken into account by imposing the
screw-boundary condition (Novotny's method).
Diagonalizing the finite-size cluster with
$N \le 20$ spins,
we observe a clear indication of continuous phase transition.
The correlation-length critical exponent is estimated
as 
$\nu=  0.92(10)$.
\end{abstract}

\pacs{
75.10.Jm        
05.30.-d        
75.40.Mg 
74.25.Ha        
}

\maketitle



\section{\label{section1}
Introduction}

According to the deconfined-criticality scenario,\cite{Senthil04,Senthil04b,Levin04}
in two dimensions,
the phase transition
separating the 
valence-bond-solid (VBS) 
and antiferromagnetic phases
is continuous;
naively,\cite{Senthil04b}
such a transition should be discontinuous,
because the adjacent phases possess distinctive
order parameters
such as the VBS coverage pattern and the sublattice magnetization,
respectively.
A good deal of theoretical investigations%
\cite{Motruich04,Tanaka06,Liu05,Dellenschneider06,Ghaemi06,Thomas07,Kim97,Singh10}
has been made to support this scenario.
(On the contrary, in Refs. 
\onlinecite{Kotov09,Isaev10,Kotov10,Isaev10b,Kuklov04,Kuklov08,Jian08,Kruger06},
it was claimed that the transition would be a weak first-order one.)

The
magnetic frustration is a clue to the realization of the VBS phase.
Actually,
the square-lattice antiferromagnet with the next-nearest-neighbor interaction
($J_1$-$J_2$ model)
exhibits the VBS phase 
around the fully frustrated ($J_2/J_1 \approx 0.5$) regime.\cite{Oitmaa96}
The quantum Monte Carlo method is not applicable to this problem
because of the negative-sign problem.
So far, 
the $J_1$-$J_2$
model has been studied with the series-expansion\cite{Oitmaa96,Sirker06} and
numerical-diagonalization\cite{Poilblanc06} methods.

Alternatively, 
one is able to realize the VBS phase
through incorporating the biquadratic interaction\cite{Read89,Kawashima07};
correspondingly, one has to enlarge the magnitude of spin to $S > 1/2$.
The biquadratic interaction (unlike the magnetic frustration)
is tractable with
the quantum Monte Carlo method.
The deconfined criticality is realized by tuning the spatial anisotropy.%
\cite{Harada07}
That is,
as the interchain interaction increases,
a transition 
from the VBS phase to either nematic or antiferromagnetic
phase occurs.\cite{Harada07}
Meanwhile,
it turned out that the ring-exchange (plaquette-four-spin) interaction 
also induces the VBS phase
even for $S=1/2$.
Extensive Monte Carlo simulations\cite{Sandvik07,Melko08}
support the deconfined-criticality
scenario; the results are compared with ours in Sec. \ref{section4}.
The transition occurs for a 
considerably large ring exchange;
the antiferromagnetic order would be far more robust than 
that of VBS.
As mentioned above,
the character of the singularity is controversial
\cite{Kotov09,Isaev10,Kotov10,Isaev10b,Kuklov04,Kuklov08,Jian08,Kruger06};
possibly, the $\log$ corrections\cite{Sandvik10} affect 
the scaling analysis.  
Even for the 
$S=1/2$ spatially anisotropic triangular antiferromagnet,\cite{Trumper99}
The ring exchange gives rise to the VBS phase.\cite{Misguich99,Nishiyama09}
This model is intractable with the quantum Monte Carlo method,
and
analytical considerations provide valuable information
as to the deconfined criticality.\cite{Xu09,Qi09,Nakane09}

\begin{figure}
\includegraphics{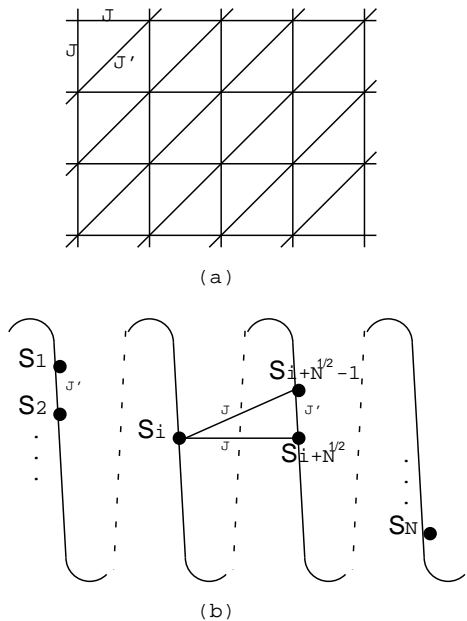}%
\caption{  \label{figure1}
(a)
We consider a spatially anisotropic triangular lattice;
the Hamiltonian is given by Eq. (\ref{Hamiltonian}).
The interaction $J$ interpolate the one- 
and two-dimensional 
lattice structures in the limiting cases $J=0$ and $J\to\infty$,
respectively.
(b)
In order to take into account such a geometrical character,
we implement the screw-boundary condition.
As shown in the drawing,
a basic structure of the cluster is
an alignment of spins $ \{ {\bf S}_i \}$
($i \le N$).
Thereby,
the dimensionality is lifted to 
$d=2$ by the bridges over the ($\sqrt{N}$)-th neighbor
pairs through the $J$ bonds.
Technical details are explicated in Sec. \ref{section2}.
}
\end{figure}

In this paper, we investigate the $S=1$ spatially-anisotropic-triangular-lattice
model with the biquadratic interaction by means of the numerical
diagonalization method.
To be specific,
the Hamiltonian is given by
\begin{equation}
\label{Hamiltonian}
{\cal H}
= 
 -J \sum_{\langle ij \rangle } 
  [ j {\bf S}_i \cdot {\bf S}_j 
         + ({\bf S}_i \cdot {\bf S}_j)^2 ]
-
J' \sum_{\langle \langle ij \rangle\rangle} 
      ({\bf S}_i \cdot {\bf S}_j)^2 
.
\end{equation}
Here,
the quantum $S=1$ spins $\{ {\bf S}_i \}$ are placed at 
each triangular-lattice point
$i$; see Fig. \ref{figure1} (a).
The summation 
$\sum_{\langle ij \rangle}$ 
($\sum_{\langle \langle ij \rangle \rangle}$)
runs over all possible nearest-neighbor
(skew-diagonal) pairs.
The parameter $J$ ($J'$) 
denotes the corresponding coupling constant.
Hereafter, we consider $J'$ as the unit of energy ($J'=1$).
Along the $J$ bond, both quadratic and biquadratic interactions 
exist, and the parameter $j$ controls a strength of the former component.
The $J'$-bond interaction is purely biquadratic.
The interaction $J$ interpolates the one-dimensional ($J=0$)
and square-lattice ($J' \to \infty$) structures.
Correspondingly, the VBS and spin-nematic phases appear,
as the interaction $J$ varies;
see a schematic phase diagram, Fig. \ref{figure2}.
In order to take into account such a geometrical character,
we implement the screw-boundary condition
[Fig. \ref{figure1} (b)] through resorting to Novotny's method
(Sec. \ref{section2}).

The rest of this paper is organized as follows.
In Sec. \ref{section2}, we explain the simulation scheme.
We also make an overview on the biquadratic-interaction spin models
relevant to ours.
In Sec. \ref{section3}, we demonstrate that the present model
exhibits a clear indication of deconfined criticality
at a moderate value of $J$.
We also analyze the criticality with the finite-size-scaling theory.
In Sec. \ref{section4}, we present the summary and discussions.

\section{\label{section2}
Screw-boundary condition: Novotny's method}

In this section,
we present the simulation scheme
(Novotny's method\cite{Novotny92,Nishiyama08}).
A brief overview on the biquadratic-interaction spin models follows.

Before commencing a explanation of technical details,
we present a basic idea of Novotny's method.
We implement the screw-boundary condition for a finite cluster
with $N$ spins; see Fig. \ref{figure1} (b).
Basically, the spins, 
$\{ {\bf S}_i \}$ ($i \le N$),
 constitute a one-dimensional ($d=1$) structure,
and the dimensionality is lifted to $d=2$ by the bridges over
the long-range pairs.
The present system 
(\ref{Hamiltonian}) has a spatial anisotropy governed by $J$.
We take into account 
such a geometrical character through imposing
the
screw-boundary condition.
According to Novotny, the long-range interactions
are
introduced systematically
by the use of the
translation operator $P$; see Eq. (\ref{HXXX}), for instance.
The operator $P$ satisfies the formula
\begin{equation}
P | S_1,S_2,\dots,S_N \rangle
   = | S_N,S_1,\dots,S_{N-1}\rangle  .
\end{equation}
Here, the base $|\{ S_i\}\rangle$ diagonalizes each of $\{ S^z_i\}$;
namely, the relation
$S^z_k | \{S_i\}\rangle = S_k | \{ S_i\}\rangle  $
holds.

Novotny's method was adapted to the quantum $S=1$ $XY$ model
in $d=2$ dimensions.\cite{Nishiyama08}
Our simulation scheme is based on this formalism.
In the following, 
we present the modifications explicitly for the sake of selfconsistency.
The $XY$ interaction $H_{XY}$, Eq. (4) of Ref. \onlinecite{Nishiyama08},
has to be replaced with the Heisenberg interaction
\begin{equation}
\label{HXXX}
H_{XXX}(v)=\sum_{i=1}^{N}(
P^vS^x_iP^{-v}S^x_i+
P^vS^y_iP^{-v}S^y_i+
P^vS^z_iP^{-v}S^z_i)    
  .
\end{equation}
Additionally, we introduce the biquadratic interaction
\begin{equation}
H_4(v)=-\frac{1}{2} H_{XXX}(v)
+\frac{1}{2} \sum_{i=1}^{N}
   \sum_{\alpha=1}^5
P^v Q^\alpha_i P^{-v} Q^\alpha_i
.
\end{equation}
The definition of $\{ Q^\alpha_i \}$ and an algebra are presented in the Appendix.
Based on these expressions,
we replace Eq. (3) of Ref. \onlinecite{Nishiyama08} with
\begin{equation}
\label{Novotny_Hamiltonian}
{\cal H}
=
-J[j H_{XXX}(\sqrt{N})+j H_{XXX}(\sqrt{N}-1)
     +H_4(\sqrt{N})+H_4(\sqrt{N}-1)]
-J' H_4(1)
.
\end{equation}
We diagonalize this matrix for $N \le 20$ spins in Sec. \ref{section3}.
The above formulae complete the formal basis of our simulation
scheme.
However,
in order to evaluate the above Hamiltonian-matrix elements
efficiently,
one may refer to a number of techniques addressed in Refs. \onlinecite{Novotny92,Nishiyama08}.


Last,
we overview the biquadratic-interaction spin models.
As mentioned in the Introduction,
the present model reduces to
the one-dimensional and square-lattice models in the limiting cases
$J=0$ and $ J \to \infty$, respectively.
Each of these limiting
cases has been studied extensively.
Here, we devote ourselves to the nearest-neighbor interaction of the
generic form,
$\cos \theta {\bf S}_i \cdot {\bf S}_j +\sin \theta ({\bf S}_i \cdot {\bf
  S}_j)^2$,
parameterized by $\theta$.
The regime $ \pi < \theta < 2\pi$ is relevant to the present research.
As for $d=1$ (Ref. \onlinecite{Fath95}),
the dimer phase appears in $ 5 \pi/4 < \theta < 7\pi/4$.
Namely, around $\theta \approx 3 \pi/2$,
the stability of
the dimer (VBS) phase would be maximal.
For $d=2$ (Ref. \onlinecite{Harada02}),
the ferromagnetic, nematic, and antiferromagnetic phases appear in 
$\theta < 5 \pi /4$,
$5 \pi /4 < \theta < 3 \pi /2$,
and 
$3 \pi /2 < \theta$, respectively.
The phase diagram in Fig. \ref{figure2} is based on these
preceding studies.

\section{\label{section3}
Numerical results}

In this section,
we present the simulation results.
We employed the simulation scheme developed
in Sec. \ref{section2}.
We devote ourselves to the analysis of the
transition between the nematic and VBS phases
(deconfined criticality); see Fig. \ref{figure2}.
We treat a variety of system sizes
$N=10,12,\dots,20$ ($N$ is the number of spins within a cluster).
The linear dimension $L$ of the cluster is
given by 
\begin{equation}
L=\sqrt{N}
  .
\end{equation}

\subsection{\label{section3_1}Critical point $J_c$}

In this section, we investigate a location of the phase boundary
separating the VBS and nematic phases.

In Fig. \ref{figure3},
we plot the scaled energy gap $L \Delta E$
for various $J$ and $N=10,12,\dots,20$.
The quadratic-interaction strength $j$ is fixed to $j=0.5$.
The symbol $\Delta E$ denotes the first excitation gap.
According to the finite-size scaling,
the scaled energy gap
$L \Delta E$ 
 should be scale-invariant at the critical point.
In fact, we observe an intersection point
at 
$J_c \approx 0.28$,
which indicates an onset of the $J$-driven phase transition.
As mentioned in the Introduction,
the nature of this singularity is under current interest.
The present result 
indicates that the singularity is a continuous one;
the critical phenomena (below the upper critical dimension)
should be described by
the finite-size scaling.

In Fig. \ref{figure4},
we plot the approximate transition point
$J_c(L_1,L_2)$
for $[2/(L_1+L_2)]^2$ with $10 \le N_1<N_2 \le 20$  
($L_{1,2} = \sqrt{N_{1,2}} $).
The parameters are the same as those of Fig. \ref{figure3}.
Here, the approximate transition point denotes a scale-invariant point with
respect to a pair of system sizes $(L_1,L_2)$.
Namely, the following relation holds;
\begin{equation}
\label{critical_point}
 L_1 \Delta E (L_1) |_{J=J_c(L_1,L_2)} = L_2 \Delta E (L_2) |_{J=J_c(L_1,L_2)}
    .
\end{equation}
The least-squares fit to the data of Fig. \ref{figure4}
yields an estimate $J_c=0.285(5)$ in the thermodynamic limit 
$L\to \infty$.
In order to appreciate possible extrapolation errors,
we made an alternative extrapolation with the $1/L^3$-abscissa scale.
Thereby, we obtain
$J_c=0.285(3)$.
The extrapolation errors appear to be negligible.
(As a matter of fact, we surveyed a wide range of $j$, and
found that the parameter $j=0.5$ yields an optimal finite-size behavior.)
Hence, we obtain an estimate
$J_c=0.285(5)$.

Making simular analyses for various values of $j$,
we arrived at a phase diagram, Fig. \ref{figure2}.
The phase boundary around $j \approx 0$ and $1$ is ambiguous
because of finite-size errors.
Possibly,
around $j \approx 1$,
the magnetic structure (antiferromagnetic order) conflicts with the 
screw-boundary condition, resulting in an enhancement of finite-size errors.
On the one hand, in $j<0$, 
the character of the transition changes to a discontinuous one,
and the finite-size-scaling method becomes invalid.
Hence, the finite-size behavior improves
around the midst ($j \approx 0.5$)
of $0 < j < 1 $.

Last, we address
a number of remarks.
First,
in the scaling analysis, Fig. \ref{figure3},
we assumed
the dynamical critical exponent $z=1$,
following the conclusion of the Monte Carlo
simulations.\cite{Sandvik07,Melko08}
Second,
we argue a possible systematic error for
the data $J_c(L_1,L_2)$ 
in Fig. \ref{figure4}.
As a matter of fact, for large system sizes,
the data $J_c(L_1,L_2)$ exhibit an enhancement,
suggesting that the extrapolated value of $J_c$
should be larger than $0.285(5)$.
However, in the subsequent analyses,
the extrapolated value of $J_c$ is no longer used,
and systematic deviations are less influential.
Nevertheless, as to the singularity of $J_c$,
it has to be mentioned that
the present data
cannot exclude a possibility of weak-first-order transition,
for which the scaling approach becomes invalidated.

\subsection{
Correlation-length critical exponent $\nu$}

In this section, we analyze the criticality 
found in Sec. \ref{section3_1}.

In Fig. \ref{figure5},
we plot the approximate critical exponent
\begin{equation}
\label{critical_exponent}
\nu(L_1,L_2)=
\frac{ \ln(L_1/L_2) }{
\ln \{ \partial_J [L_1 \Delta E(L_1)]/\partial_J [L_2\Delta E(L_2)]\}|_{J=J_c(L_1,L_2)}  }
           ,
\end{equation}
for $2/(L_1+L_2)$ with 
$10 \le N_1<N_2 \le 20$.
The parameters are the same as those of Fig. \ref{figure3}.
The least-squares fit to these data yields $\nu=0.97(2)$
in the thermodynamic limit.
This estimate may be affected by systematic errors.
(The statistical error would be an underestimate.)
In order to appreciate an error margin,
we made an alternative extrapolation with the $1/L^2$-abscissa scale.
Thereby, we obtain
$\nu=0.86(1)$.
The discrepancy indicates an amount of systematic errors.
As a result,
we  arrive at an estimate
\begin{equation}
\nu=   0.92(10)   ,
\end{equation}
 which covers the above results obtained via independent extrapolations.

This is a good position to address a remark on the scaling behaviors of
Figs. \ref{figure3}-\ref{figure5}.
As mentioned in the Introduction, notorious $\log$ corrections
were observed for the $S=1/2$ square-lattice antiferromagnet with the ring
exchange.
Our data, on the contrary, appear to exhibit moderate corrections to scaling,
particularly, in Figs. \ref{figure3} and \ref{figure4}.
For the former antiferromagnet, the VBS phase emerges for a considerably large
ring exchange, indicating that the antiferromagnetic phase is robust.
On the one hand, our model exhibits a stable VBS state owing to the spatial
anisotropy (one-dimensionality in the $J=0$ limit),
and the nematic phase turns into the VBS phase 
at a moderate coupling strength, $J_c \approx 0.3$.
We suspect that these peculiarities of the present (rather artificially designed) model
bring about improved scaling behaviors.

Last, we make a consideration of the 
abscissa scale of Fig. \ref{figure5} (\ref{figure4}).
The critical exponent
(point) has a leading correction of O$(L^{-\omega})$
[O$(L^{-\omega-1/\nu})$];
here, the symbol $\omega$ denotes 
the index for corrections to scaling.
At present, the index $\omega$ for the deconfined criticality
is unclear.
As a reference, one may refer to that of
the $d=3$ Heisenberg universality class,
$\omega=0.773$ (Ref. \onlinecite{Hasenbusch01}).
Making use of this value, we set the abscissa scale 
to that depicted in
Fig. \ref{figure5} (\ref{figure4}).

\section{\label{section4}
Summary and discussions}

\begin{figure}
\includegraphics{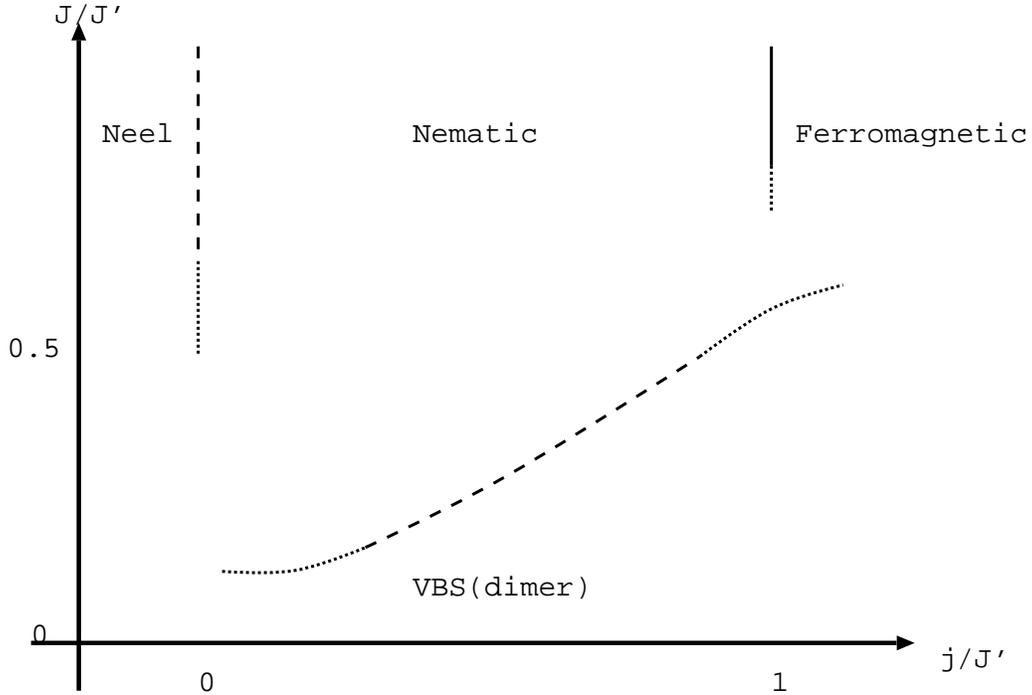}%
\caption{  \label{figure2}
A schematic phase diagram for the 
$S=1$
spatially-anisotropic-triangular-lattice model, Eq. 
(\ref{Hamiltonian}),
is presented.
The limiting cases $J \to 0$ and $ J \to \infty$
were studied in Refs. \onlinecite{Fath95} and \onlinecite{Harada02},
respectively.
The solid (dashed) lines stand for the phase boundaries
of discontinuous (continuous) character.
The dotted lines are ambiguous.
We investigate the phase boundary separating the nematic and
VBS phases.
}
\end{figure}

The $S=1$ spin model on the spatially anisotropic triangular lattice, Eq. (\ref{Hamiltonian}),
was investigated numerically.
As the spatial anisotropy $J$ changes,
the VBS and nematic phases appear
(Fig. \ref{figure2}).
Hence, this rather artificial model provides a candidate for the analysis of the
deconfined criticality, which is arousing much attention recently.
We employed Novotny's method (screw boundary condition)
in order to take into account such a geometrical character
(spatial anisotropy).

\begin{figure}
\includegraphics{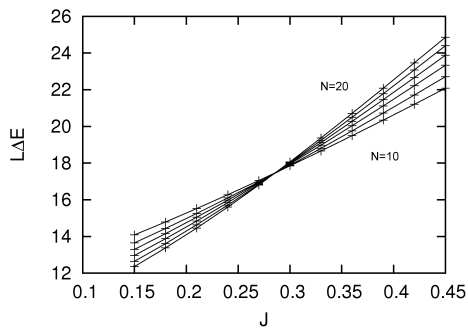}%
\caption{  \label{figure3}
The scaled energy gap $L \Delta E$
is plotted for various $J$ and $N=10,12,\dots,20$.
The quadratic-interaction strength $j$ is fixed to $j=0.5$.
($J'$ is the unit of energy.)
We observe a clear indication of the deconfined criticality
around $J \approx 0.28$.
}
\end{figure}

\begin{figure}
\includegraphics{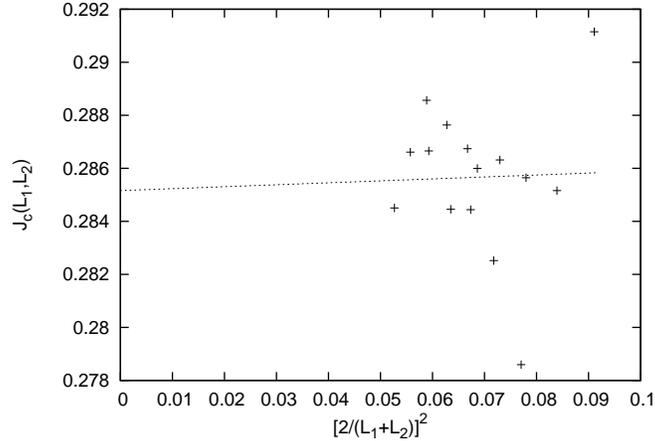}%
\caption{  \label{figure4}
The approximate critical point $J_c(L_1,L_2)$ 
(\ref{critical_point}) is plotted
for $[2/(L_1+L_2)]^2$ with 
$10 \le N_1 <N_2 \le 20$.
The parameters are the same as those of Fig. \ref{figure3}. 
The least-squares fit to these data yields 
$J_c=0.285(5)$
in the thermodynamic limit.
}
\end{figure}

\begin{figure}
\includegraphics{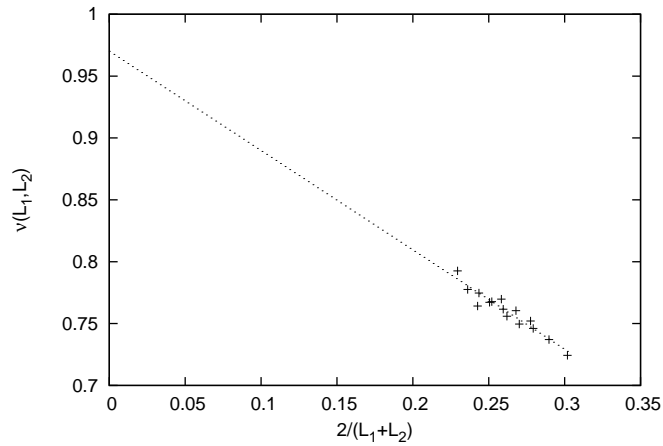}%
\caption{  \label{figure5}
The approximate critical exponent $\nu(L_1,L_2)$ (\ref{critical_exponent})
is plotted for $2/(L_1+L_2)$ with $10 \le N_1<N_2 \le 20$.
The parameters are the same as those of Fig. \ref{figure3}.
 The least-squares fit to these data yields
$\nu=0.97(2)$
in the thermodynamic limit.
A possible systematic error is considered in the text.
}
\end{figure}

As a result, we observe a clear indication of the $J$-driven criticality
through the finite-size-scaling analysis (Fig. \ref{figure3}).
This result supports the deconfined-criticality scenario.
Thereby, we estimate the
correlation-length critical exponent
as $\nu=0.92(10)$.

As a reference,
we overview related studies.
For the square-lattice antiferromagnet with the ring exchange,
the
estimates, $\nu=0.78(3)$ (Ref. \onlinecite{Sandvik07}) and 
$\nu=0.68(4)$ (Ref. \onlinecite{Melko08}), were reported.
As for the spatially-anisotropic-triangular
antiferromagnet
with the ring exchange, the exponent $\nu=0.80(15)$ was obtained.\cite{Nishiyama09}
These results are to be compared with the index,
$\nu=0.7112(5)$ (Ref. \onlinecite{Campostrini02}),
for 
the $d=3$ Heisenberg universality class.
Our result indicates a tendency toward an enhancement for the correlation-length critical exponent,
as compared to that for the $d=3$ Heisenberg universality class.
Taking the advantage of the numerical diagonalization method,
we are able to extend the interactions so as to eliminate
finite-size errors.
A frustrated interaction along the $J'$-bond direction may 
stabilize
(extend the regime of)
the VBS phase substantially.
This problem will be addressed in the
future study.


%

%



\appendix*
\section{\label{appendix}
A reduction formula for the biquadratic interaction}

The biquadratic interaction 
$({\bf S}_i \cdot {\bf S}_j)^2$ 
reduces to a seemingly quadratic form
\begin{equation}
({\bf S}_i \cdot {\bf S}_j)^2 =
  -{\bf S}_i \cdot {\bf S}_j/2
  +\sum_{\alpha=1}^{5} Q_i^\alpha Q_j^\alpha/2+4/3
.
\end{equation}
Here, the operators $\{ Q_i^\alpha \}$
are given by the relations,
$Q_i^1 = (S_i^x)^2-(S_i^y)^2$,
$Q_i^2 = [2(S_i^z)^2-(S_i^x)^2-(S_i^y)^2]/ \sqrt{3}$,
$Q_i^3 = S_i^xS_i^y+S_i^yS_i^x$,
$Q_i^4 = S_i^yS_i^z+S_i^zS_i^y$,
and
$Q_i^5 = S_i^xS_i^z+S_i^zS_i^x$.
This reduction formula is a key ingredient in Sec. \ref{section2}.



\end{document}